\documentclass[conference]{IEEEtran}

\usepackage{amsmath,amssymb,amsfonts}
\usepackage{balance}
\usepackage{cite}
\usepackage{enumitem}
\usepackage{graphicx}
\usepackage{textcomp}
\usepackage{tikz}
\usepackage{url}
\usepackage{xcolor}
\usetikzlibrary{arrows.meta,positioning}

\def\BibTeX{{\rm B\kern-.05em{\sc i\kern-.025em b}\kern-.08em
    T\kern-.1667em\lower.7ex\hbox{E}\kern-.125emX}}

\newcommand{\stitle}[1]{{\noindent\bfseries #1}}
\newcommand{\sstitle}[1]{%
  \ifhmode%
    {\textit{\underline{#1}}}%
  \else%
    {\noindent\textit{\underline{#1}}}%
  \fi}
\newcommand{\ssstitle}[1]{%
  \ifhmode%
    {\textit{#1}}%
  \else%
    {\noindent\textit{#1}}%
  \fi}
\newcommand{\authorcolwidth}{0.245\textwidth}
\makeatletter
\newcommand{\authorcol}[3]{%
  \begin{minipage}[t]{\authorcolwidth}
    \centering
    {\@IEEEauthorblockNstyle #1\par}
    {\@IEEEauthorblockAstyle \itshape #2\\\upshape #3\par}
  \end{minipage}}
\newcommand{\authorcolnowrap}[3]{%
  \begin{minipage}[t]{\authorcolwidth}
    \centering
    {\@IEEEauthorblockNstyle #1\par}
    {\@IEEEauthorblockAstyle \itshape \makebox[\authorcolwidth][c]{#2}\\\upshape #3\par}
  \end{minipage}}
\makeatother
\newcommand{\contractterm}[1]{\textit{#1}}

\makeatletter
\def\section{\@startsection{section}{1}{\z@}{1.0ex plus 0.8ex minus 0.3ex}%
{0.45ex plus 0.5ex minus 0ex}{\normalfont\normalsize\centering\scshape}}%
\makeatother

\title{RAIDS: Rethinking Data Systems as Responsible Intelligent Infrastructure}

\author{%
  \begin{tabular}{@{}c@{}c@{}c@{}c@{}}
    \authorcol{Zhengyi Yang}
    {University of Sydney}
    {zhengyi.yang@sydney.edu.au}
    &
    \authorcol{Wenke Yang}
    {University of New South Wales}
    {wenke.yang@unsw.edu.au}
    &
    \authorcol{Guanfeng Liu}
    {Macquarie University}
    {guanfeng.liu@mq.edu.au}
    &
    \authorcolnowrap{Lu Qin}
    {University of Technology Sydney}
    {lu.qin@uts.edu.au}
  \end{tabular}\vspace{-0.3em}
}

\begin{document}
\maketitle

\begin{abstract}
Data systems are evolving from information infrastructure into decision infrastructure. Yet responsibility mechanisms have not kept pace: an output can be accurate or efficient while still lacking sufficient support, satisfied constraints, and actionability for responsible use.
We propose RAIDS (\textbf{R}esponsible \textbf{a}nd \textbf{I}ntelligent \textbf{D}ata \textbf{S}ystem), a vision for data systems as responsible intelligent infrastructure.
RAIDS treats responsibility not as post-hoc metadata, but as \emph{execution semantics} for holistic data-to-decision and data mining pipelines.
Its core abstraction is an operator-level \emph{responsibility contract}: each operator exposes an output together with \emph{support}, \emph{constraint}, and \emph{actionability} state under an explicit responsibility context, and these contracts compose across pipelines.
These states capture whether an output is grounded, whether execution satisfies relevant limits, and which action modes are permissible.
We introduce \emph{responsibility preservation} as the organizing systems objective: responsibility state should remain sufficient as execution proceeds, or the system should repair, replan, escalate, refuse, or otherwise change course.
We outline a BlueSky research agenda for RAIDS, spanning responsibility-preserving execution, responsibility-aware optimization, provenance, oversight, and evaluation.
\end{abstract}


\section{Introduction}

Data systems are evolving from information infrastructure into decision infrastructure.
Beyond storing, querying, and analyzing data, they are increasingly used to support real-world decision making across critical domains including health care~\cite{ObermeyerScience2019Bias,SinghalNature2023ClinicalKnowledge,yan2025salpcg}, finance~\cite{GuJFE2020AssetPricing,wang2025aefa,shu2026forexagent}, public services~\cite{EngstromACUS2020GovernmentAlgorithm,KrollUPennLR2017AccountableAlgorithms,ChouldechovaBigData2017FairPrediction,EnsignFAT2018RunawayFeedback,yan2026improving}, scientific discovery~\cite{Hey2009FourthParadigm,JumperNature2021AlphaFold,MerchantNature2023MaterialsDiscovery}, climate response~\cite{BiNature2023PanguWeather,HinoNatSustain2018EnvMonitoring,chen2023machine}, and AI-assisted work~\cite{LiangTMLR2023HELM,tang2025tabular,yang2024parallel}.
This transformation spans the entire data intelligence stack, from database systems~\cite{JagadishCACM2014BigDataChallenges,AilamakiArxiv2025CambridgeReport,PanVLDBJ2024VectorSurvey} and data mining~\cite{JordanScience2015MachineLearning,DataPerf2022,CortesVLDBW2024DataQualityResponsibleAI,NingICDM2025TimeSeriesReasoning} to emerging LLM-powered data analytics and applications~\cite{GaoPVLDB2024Text2SQL,PatelPVLDB2025SemanticOperators,ShankarArxiv2024DocETL,lai2025graphy}.

\stitle{Capability requires responsibility.}
As data systems become embedded in real-world decision making, their outputs carry consequences beyond information access and analysis.
In critical domains, system outputs can influence decisions about care, credit, public resources, scientific priorities, environmental response, and everyday work~\cite{NIST2023AIRMF}.
A pipeline can no longer be judged solely by efficiency, scalability, accuracy, or utility; it must also be judged by whether its outputs are appropriate for use in context.
A technically strong result may still be unsuitable when it is unsupported by sufficient evidence, violates applicable constraints or regulations, imposes unnecessary environmental costs, or requires human judgment before action.
Together, these developments point to a broader question for the community: \textit{how should systems determine whether an output is responsible enough to use?}

\stitle{The responsibility gap.}
Emerging efforts have raised responsibility-related concerns in data systems, including fairness and transparency~\cite{StoyanovichCACM2022ResponsibleDM,AbiteboulArxiv2019TransparencyFairness}, governance and accountability~\cite{KhatriCACM2010DesigningDataGovernance,RajiFAccT2020AccountabilityGap}, privacy and data protection~\cite{DworkTCC2006DifferentialPrivacy,EuropeanUnion2016GDPR}, sustainability~\cite{BachrasPVLDB2025SustainableDB}, safety, and human oversight~\cite{NIST2023AIRMF,EuropeanUnion2024AIAct}.
Existing systems also expose fragments of responsibility through mechanisms such as data validation~\cite{BreckMLSys2019DataValidation}, documentation~\cite{MitchellFAccT2019ModelCards,GebruCACM2021Datasheets}, provenance~\cite{GreenPODS2007ProvenanceSemirings,PsallidasPVLDB2023OneProvenance}, and auditing~\cite{Sandvig2014AuditingAlgorithms,RajiFAccT2020AccountabilityGap,RajiAIES2019ActionableAuditing}.
These mechanisms provide necessary visibility, but they often \emph{sit beside execution}: their outputs are recorded, documented, or checked without participating in runtime decisions, control flow, or action protocols.
As outputs move across operators, models, teams, and deployment boundaries, downstream actors must reconstruct trust, constraints, uncertainty, and appropriate use after the fact.
Evidence may be lost, policy constraints may be dropped, environmental impacts may remain invisible, and human review may never be triggered.
The responsibility gap is not the absence of another checklist or governance component; it is a systems gap: \textit{responsibility is not yet preserved and acted on throughout the data-to-decision process.}

\stitle{Why now.}
Three shifts make this question timely.
Data mining, AI analytics, and LLM-empowered pipelines now turn query results and mined patterns directly into claims, recommendations, and actions.
At the same time, privacy, fairness, safety, sustainability, auditability, and human oversight increasingly determine whether an output can be used, not merely documented.
Modern data infrastructure already exposes the needed hooks: metadata, validation, provenance, policy checks, monitoring, optimization, and control flow.
The BlueSky question is whether these hooks can become execution semantics: \textit{can responsibility change what a pipeline computes, preserves, repairs, escalates, or refuses?}

\stitle{The RAIDS idea.}
We propose RAIDS (\textbf{R}esponsible \textbf{a}nd \textbf{I}ntelligent \textbf{D}ata \textbf{S}ystem), a vision for responsibility-native data systems.
The central claim is \emph{responsibility as execution semantics}: responsibility should participate in execution rather than remain external to it.
A data system should expose not only an output, but whether it has sufficient \emph{support}, satisfies relevant \emph{constraints}, and has appropriate \emph{actionability} in context.
RAIDS asks which pipelines are admissible, which plans are preferable, what evidence must be preserved, when outputs should be qualified, and when execution should repair, escalate, refuse, audit, invite review, or stop execution.
Responsibility is not an after-the-fact assessment but a first-class systems concern throughout the data-to-decision process.

\stitle{Contributions.}
The paper makes three contributions. Together, they recast responsibility as a systems problem and define a research agenda for responsibility-native data systems.

\begin{itemize}[wide,noitemsep,topsep=0pt,parsep=0pt,partopsep=0pt]

    \item \stitle{Responsibility-native infrastructure.}
    We argue that as data systems evolve from information infrastructure into decision infrastructure, responsibility must become a native systems concern. We recast modern data systems as a \emph{Responsible Data Intelligence Loop}, in which responsibility is preserved throughout the data-to-decision process.

    \item \stitle{Executable responsibility contract.}
    We introduce a responsibility contract as a systems primitive for \emph{responsibility as execution semantics}. The contract exposes \emph{support}, \emph{constraint}, and \emph{actionability} under an explicit responsibility context, enabling responsibility to be propagated, optimized, monitored, audited, and used to govern execution.

    \item \stitle{A BlueSky research agenda.}
    We outline a research agenda organized around \emph{responsibility preservation}: sufficient support, satisfied constraints, and appropriate actionability throughout a data-to-decision pipeline. The agenda connects execution, optimization, provenance, oversight, and evaluation within a unified systems perspective.

\end{itemize}

\section{Limits of Current Approaches}

\stitle{From visibility to execution.}
Prior work provides building blocks for RAIDS, but the dominant pattern remains visibility rather than execution.
Capability-oriented systems make data infrastructure more automated, adaptive, and model-mediated: deployed data and AI pipelines expose lifecycle debt~\cite{PolyzotisSIGMOD2017ProductionML,SculleyNeurIPS2015TechnicalDebt}; self-driving databases automate system decisions~\cite{PavloPVLDB2021ElectricSheep}; and LLM-facing interfaces~\cite{GaoPVLDB2024Text2SQL}, semantic operators~\cite{PatelPVLDB2025SemanticOperators}, and vector substrates~\cite{PanVLDBJ2024VectorSurvey} move retrieval, interpretation, and generation into data management.
These advances expand capability; RAIDS asks \emph{what responsibility state must accompany outputs as they become inputs to action?}

\stitle{Foundational but fragmented.}
Responsibility-oriented work exposes information for responsible use.
Responsible data management frames fairness, transparency, governance, and accountability as data-system concerns~\cite{StoyanovichCACM2022ResponsibleDM}; data governance identifies decision rights, responsibilities, and context-dependent coordination mechanisms~\cite{KhatriCACM2010DesigningDataGovernance,WeberJDIQ2009ContingencyDataGovernance,AbrahamIJIM2019DataGovernanceFramework,JanssenGIQ2020DataGovernanceTrustworthyAI}.
Provenance captures derivations and traces~\cite{GreenPODS2007ProvenanceSemirings,ChapmanPVLDB2021PreprocessingProvenance,PsallidasPVLDB2023OneProvenance}; auditing and influence analysis expose accountability gaps and input effects~\cite{RajiFAccT2020AccountabilityGap,DattaSP2016InputInfluence}; documentation records intended use and limitations~\cite{MitchellFAccT2019ModelCards,GebruCACM2021Datasheets}; and machine-readable metadata supports responsible-use fields~\cite{AkhtarNeurIPS2024Croissant,JainArxiv2024CroissantRAI}.
Fairness-aware repair~\cite{SalimiSIGMOD2019InterventionalFairness}, integration~\cite{NargesianPVLDB2021FairDI}, and ranking~\cite{SinghKDD2018FairExposure,BiegaSIGIR2018EquityAttention} expose social constraints; sustainability work exposes environmental costs~\cite{BachrasPVLDB2025SustainableDB}; validation and repair expose quality failures~\cite{BreckMLSys2019DataValidation,RekatsinasPVLDB2017HoloClean,KrishnanSIGMOD2016ActiveClean}; label-quality analysis exposes annotation failures~\cite{NorthcuttJAIR2021ConfidentLearning}; and retrieval-augmented generation (RAG) evaluation exposes evidence-use failures~\cite{YangArxiv2024CRAG}.
Together, these streams make responsibility visible, documented, explainable, and auditable, but they remain disconnected fragments. RAIDS asks when this information should become part of execution itself.

\stitle{The missing coordination.}
Current approaches lack end-to-end coordination, not responsibility information.
Provenance can explain data origins; fairness and sustainability mechanisms can reveal constraints; documentation can expose assumptions; and RAG evaluation can reveal unsupported generation.
Yet these artifacts often remain external to plan selection, retrieval steering, ranking objectives, action protocols, escalation policies, and refusal decisions.
RAIDS asks when responsibility information becomes execution state that invalidates plans, requires repair, changes rankings, triggers review, qualifies outputs, escalates decisions, or stops automation.

\section{The Responsible Data Intelligence Loop}

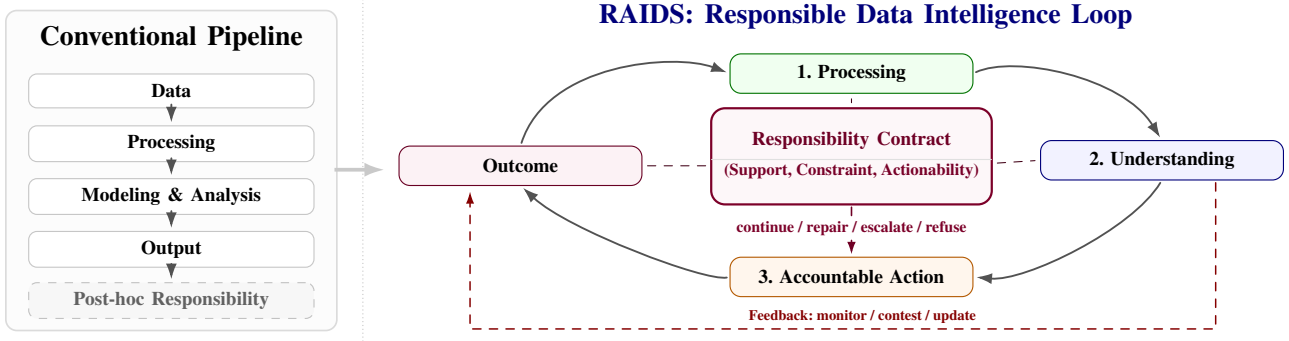
\begin{figure*}[t]

\centering
\resizebox{0.94\textwidth}{!}{%
  \begin{tikzpicture}[
      every node/.style={outer sep=0pt},
      panel/.style={
          draw=gray!28,
          rounded corners=4pt,
          fill=gray!2,
          line width=0.6pt
        },
      convbox/.style={
          draw=gray!45,
          rounded corners=3pt,
          fill=white,
          minimum width=3.62cm,
          minimum height=0.43cm,
          align=center,
          inner xsep=6pt,
          font=\scriptsize\bfseries
        },
      convcheck/.style={
          draw=gray!48,
          dashed,
          rounded corners=3pt,
          fill=gray!6,
          minimum width=3.62cm,
          minimum height=0.44cm,
          align=center,
          inner xsep=6pt,
          font=\scriptsize\bfseries,
          text=black!62
        },
      loopnode/.style={
          rounded corners=4pt,
          minimum width=2.90cm,
          minimum height=0.50cm,
          align=center,
          text width=2.66cm,
          inner xsep=6pt,
          inner ysep=3pt,
          font=\scriptsize\bfseries
        },
      processing/.style={
          loopnode,
          draw=green!45!black,
          fill=green!6
        },
      understanding/.style={
          loopnode,
          draw=blue!48!black,
          fill=blue!5
        },
      actionbox/.style={
          loopnode,
          draw=orange!70!black,
          fill=orange!7
        },
      outcomebox/.style={
          loopnode,
          draw=purple!55!black,
          fill=purple!5
        },
      contract/.style={
          draw=purple!65!black,
          rounded corners=5pt,
          fill=purple!3,
          line width=0.8pt
        },
      state/.style={
          align=center,
          font=\fontsize{6}{6.5}\selectfont\bfseries,
          inner xsep=6pt,
          inner ysep=2pt,
          text=purple!63!black
        },
      condition/.style={
          fill=white,
          align=center,
          font=\fontsize{5.7}{6.1}\selectfont\bfseries,
          inner xsep=4pt,
          inner ysep=1pt,
          text=purple!65!black
        },
      flow/.style={
          -{Latex[length=2.0mm,width=1.4mm]},
          line width=0.68pt,
          draw=black!68,
          shorten >=1.2pt,
          shorten <=1.2pt,
          line cap=round,
          line join=round
        },
      contractlink/.style={
          -{Latex[length=1.6mm,width=1.1mm]},
          dashed,
          line width=0.58pt,
          draw=purple!58!black,
          shorten >=1pt,
          shorten <=1pt,
          line cap=round
        },
      contractaux/.style={
          dashed,
          line width=0.38pt,
          draw=purple!34!black,
          shorten >=2pt,
          shorten <=2pt,
          line cap=round
        },
      feedback/.style={
          -{Latex[length=1.55mm,width=1.05mm]},
          dashed,
          line width=0.54pt,
          draw=red!52!black,
          shorten >=1pt,
          shorten <=1pt,
          line cap=round
        }
    ]

    % Conventional pipeline.
    \draw[panel] (0.05,-2.04) rectangle (4.25,2.02);
    \node[anchor=center,font=\bfseries\normalsize] at (2.15,1.67) {Conventional Pipeline};

    \node[convbox] (cdata) at (2.15,1.00) {Data};
    \node[convbox] (cproc) at (2.15,0.33) {Processing};
    \node[convbox] (cmodel) at (2.15,-0.34) {Modeling \& Analysis};
    \node[convbox] (cout) at (2.15,-1.01) {Output};
    \node[convcheck] (cpost) at (2.15,-1.66) {Post-hoc Responsibility};

    \draw[flow] (cdata.south) -- (cproc.north);
    \draw[flow] (cproc.south) -- (cmodel.north);
    \draw[flow] (cmodel.south) -- (cout.north);
    \draw[flow] (cout.south) -- (cpost.north);

    % Transition divider.
    \draw[densely dotted,gray!45] (4.58,-2.18) -- (4.58,2.18);
    \draw[-{Latex[length=2.6mm,width=1.8mm]},line width=1.35pt,draw=gray!42] (4.22,0.00) -- (4.86,0.00);

    % RAIDS loop.
    \node[anchor=north,font=\bfseries\normalsize,text=blue!48!black] at (10.96,2.22)
      {RAIDS: Responsible Data Intelligence Loop};

    \node[processing] (processing) at (10.78,1.22)
      {1. Processing};
    \node[understanding] (understanding) at (14.72,0.12)
      {2. Understanding};
    \node[actionbox] (action) at (10.78,-1.36)
      {3. Accountable Action};
    \node[outcomebox] (outcome) at (6.58,0.05)
      {Outcome};

    \draw[flow]
      (outcome.north) .. controls (6.92,1.36) and (8.54,1.48) .. (processing.west);
    \draw[flow]
      (processing.east) .. controls (12.84,1.46) and (14.22,1.18) .. (understanding.north);
    \draw[flow]
      (understanding.south) .. controls (14.34,-0.76) and (12.98,-1.42) .. (action.east);
    \draw[flow]
      (action.west) .. controls (8.92,-1.42) and (7.18,-0.76) .. (outcome.south);

    % Central responsibility contract.
    \draw[contract] (9.00,-0.44) rectangle (12.56,0.78);
    \node[
      anchor=north,
      font=\scriptsize\bfseries,
      text=purple!58!black,
      align=center
    ] at (10.78,0.60) {Responsibility Contract};

    \draw[gray!24] (9.00,0.13) -- (12.56,0.13);

    \node[state] at (10.78,-0.01) {(Support, Constraint, Actionability)};

    \draw[contractaux] (10.78,0.78) -- (processing.south);
    \draw[contractaux] (12.56,0.08) -- (understanding.west);
    \draw[contractlink] (10.78,-0.44) --
      node[condition,pos=0.45] {continue / repair / escalate / refuse}
      (action.north);
    \draw[contractaux] (9.00,0.05) -- (outcome.east);

    \draw[feedback]
      ([xshift=-0.86cm]understanding.south east) |- (5.94,-2.02)
      -- (5.94,-0.24);
    \node[
      anchor=center,
      align=center,
      font=\fontsize{5.25}{5.8}\selectfont\bfseries,
      text=red!52!black,
      fill=white,
      inner xsep=3pt,
      inner ysep=0.8pt
    ] at (10.92,-1.86) {Feedback: monitor / contest / update};

  \end{tikzpicture}
}
\caption{Conventional pipelines attach responsibility after output. RAIDS makes the responsibility contract part of the loop: responsibility state (support, constraint, and actionability) travels through processing, understanding, accountable action, and feedback before returning to outcome.}
\label{fig:loop}
\vspace{-0.45em}

\end{figure*}

\stitle{From pipeline to loop.}
Conventional data pipelines assume that execution ends when an output is produced, so responsibility is assessed at the point of output.
Modern data systems increasingly operate inside decision cycles: outputs become inputs to action, and consequences return through feedback to influence future data, policies, and decisions.
When support is insufficient, constraints are violated, or actionability is uncertain, execution should not simply continue; systems may repair evidence, replan, request review, escalate, or refuse unsafe action before re-entering the loop.
Responsibility therefore does not terminate at output; the needed abstraction is a loop, not a one-way pipeline.
RAIDS views data systems through the \emph{Responsible Data Intelligence Loop} in Fig.~\ref{fig:loop}, where responsibility must be preserved across processing, understanding, accountable action, and feedback.

\stitle{Responsibility as loop state.}
Responsibility is not a layer outside the loop; it is loop state: what supports the output, which constraints apply, and which action should follow.
If this state is insufficient, systems should repair it, qualify the result, escalate, refuse, or record audit conditions.
Responsibility preservation means preserving this state when sufficient, repairing it when degraded, and using it to change course when insufficient.
This is the visual claim in Fig.~\ref{fig:loop}: RAIDS puts \emph{responsibility in the loop}; the contract links loop stages, the solid flow routes understanding through accountable action, and the dashed curve marks feedback: monitor, contest, or update outcome state before re-entry.

\sstitle{Loop phases.}
In \emph{processing}, systems identify relevant data, preserved quality evidence, affected groups, applicable policies, and costs.
In \emph{understanding}, systems turn data into claims, uncertainty, explanations, and provenance.
In \emph{accountable action}, systems decide whether the result should be used, qualified, reviewed, refused, audited, or improved.
Responsibility state couples these phases: weak processing can make claims unsupported, weak understanding can hide uncertainty, and weak action protocols can make an output irresponsible.

\sstitle{Scope of the vision.}
RAIDS is neither post-training responsible AI nor a governance layer.
Failures arise before model invocation; post-hoc governance cannot change execution.
Nor is RAIDS reducible to provenance, fairness, sustainability, or explainability, each of which captures a fragment.
The problem is responsibility coordination: how should infrastructure preserve, repair, and operationalize responsibility state as pipelines move from data to decision and feedback?

\stitle{The BlueSky opportunity.}
The opportunity is to make responsibility executable.
Database research has turned broad goals into systems abstractions: consistency into transaction semantics, performance into optimization, reliability into recovery, and scale into distributed execution.
RAIDS asks for an analogous move in responsible data intelligence.
The Responsible Data Intelligence Loop defines the system boundary; the responsibility contract defines the execution interface through which responsibility participates in computation.

\section{The Responsibility Contract}

Making responsibility native requires an infrastructure primitive: a \emph{responsibility contract}. This is the core abstraction: the interface through which responsibility participates in execution and tests whether operators preserve it when they compose. A conventional operator contract specifies inputs, outputs, cost, and accuracy; a RAIDS contract exposes epistemic support, normative constraints, and allowable action modes.

\begin{center}
  \begin{tabular}{@{}l@{}}
    \textit{DataOperator}(\contractterm{input}, \contractterm{task\_context}, \contractterm{responsibility\_context})\\
    $\rightarrow$ (\contractterm{output}, \contractterm{support}, \contractterm{constraint}, \contractterm{actionability}).
  \end{tabular}
\end{center}

\stitle{Context.} Here \emph{input} denotes input data and \emph{output} the produced result. The \emph{task\_context} describes task type, intent, scope, target output, domain, time window, and pipeline role. The \emph{responsibility\_context} specifies obligations and thresholds for judging the output; the same mining, retrieval, ranking, or decision-support task may be exploratory in one context and regulated in another. RAIDS uses four compact dimensions:

\begin{itemize}[wide,noitemsep,topsep=0pt,parsep=0pt,partopsep=0pt]
  \item \ssstitle{Evidence and quality}: evidence, metadata, uncertainty, provenance, freshness, robustness, and explanation~\cite{WilkinsonSciData2016FAIR,LiICDM2023DataQualityTreatment,DoddaiahICDM2022TimeSeriesExplainability}.
  \item \ssstitle{Policy and accountability}: regulation~\cite{EuropeanUnion2016GDPR,EuropeanUnion2024AIAct}, governance decision rights~\cite{KhatriCACM2010DesigningDataGovernance}, accountable roles, privacy duties, reporting duties, and audit requirements that parameterize execution.
  \item \ssstitle{Social and environmental impact}: affected groups and distributional effects~\cite{HardtNeurIPS2016EqualOpportunity,HuICDM2020MetricFreeIndividualFairness,ZhangICDM2021FairUncertainty,SinghKDD2018FairExposure,tang2025understand}, sustainability budgets and footprint~\cite{BachrasPVLDB2025SustainableDB}, and social-benefit obligations~\cite{UnitedNations2015SDGs}.
  \item \ssstitle{Decision and oversight}: qualification rules, human-in-the-loop triggers, review, escalation, refusal, contestability, and feedback~\cite{WhiteHouse2022AIBillRights,NIST2023AIRMF}.
\end{itemize}

\noindent These dimensions keep responsibility contextual without turning RAIDS into a compliance framework. Data-management work links transparency, fairness, and data protection~\cite{AbiteboulArxiv2019TransparencyFairness}; privacy mechanisms and regulation shape disclosure and data use~\cite{DworkTCC2006DifferentialPrivacy,FeyisetanICDM2019PrivacyUtilityText,EuropeanUnion2016GDPR}; and policy frameworks emphasize human fallback, notice, explanation, safety, and discrimination protection~\cite{WhiteHouse2022AIBillRights}. In RAIDS, these obligations become execution-changing parameters.

\sstitle{Executable states.} The contract has three validity states.

\begin{itemize}[wide,noitemsep,topsep=0pt,parsep=0pt,partopsep=0pt]
  \item \ssstitle{Support} provides epistemic validity: whether the output is sufficiently grounded by evidence, freshness, uncertainty, provenance, conflicts, quality, and explanation.
  \item \ssstitle{Constraint} provides normative validity: whether the output and execution path satisfy policy, fairness, privacy, safety, sustainability, resource, and impact limits.
  \item \ssstitle{Actionability} provides operational validity: what the next protocol should be, such as answer, qualify, warn, ask, defer, escalate, refuse, audit, or report.
\end{itemize}

\noindent Together, support, constraint, and actionability correspond to epistemic, normative, and operational validity. Constraint state is executable because it determines which outputs, operators, and continuations remain admissible. For example, fairness criteria~\cite{DworkITCS2012FairnessAwareness,HardtNeurIPS2016EqualOpportunity,ZhangICDM2021FairUncertainty} and privacy requirements~\cite{DworkTCC2006DifferentialPrivacy,ChenICDM2025FairLDP,EuropeanUnion2016GDPR} are not labels attached after execution; they can change how data are selected, repaired, aggregated, anonymized, ranked, withheld, qualified, or refused before use.

\noindent The states guide continuation. If insufficient, RAIDS may refresh evidence, preserve conflict, replan, reduce footprint, ask, escalate, refuse, invite contestation, or produce an audit trail. The goal is to expose enough state for responsible use, not to make every component solve every social objective: delayed harms can arise when actions influence future data~\cite{LiuICML2018DelayedImpact,EnsignFAT2018RunawayFeedback}.

\sstitle{Systems implications.} The contract implies four system behaviors: preserve or repair responsibility state across operators; use that state in optimization; keep it inspectable for evidence, policy, impact, review, and risk; and let weak support, violated constraints, or uncertain actionability alter control flow through qualification, clarification, escalation, or refusal.

\stitle{Responsibility as execution semantics.} Here responsibility becomes execution semantics rather than decoration. A responsibility contract changes the set of valid pipelines, not just the explanation shown after execution. It can make an optimizer prefer stronger evidence, refresh stale sources, preserve conflict, or require human review before action. In traditional analytics, the system boundary ends at a query result, dashboard, or prediction; in responsible data intelligence, it extends to the conditions under which the output becomes actionable. A result that cannot be explained, audited, or safely used is incomplete, even if syntactically valid and statistically accurate; RAIDS treats refusal, qualification, escalation, contestation, and uncertainty exposure as system behaviors.

\balance
\section{A BlueSky Research Agenda}

\noindent The target is \emph{responsibility preservation}: after composition, the system should still determine whether an output is supported, constraint-compliant, and actionable. If state degrades, execution should repair it or change course. Responsibility state remains visible, compositional, optimizable, and actionable.

\stitle{Responsibility state management.} RAIDS needs a state model for support, constraint, and actionability. The challenge is to represent heterogeneous responsibility information in a form lightweight enough for execution yet structured enough to be propagated, composed, degraded, and repaired. The state should carry evidence, uncertainty, provenance, quality, policy obligations, fairness constraints, sustainability limits, review requirements, and unresolved risks. For example, in evidence-intensive pipelines such as RAG~\cite{LewisNeurIPS2020RAG,EdgeArxiv2024GraphRAG,liu2026a2rag}, retrieved passages, provenance, freshness, coverage, conflicts, claim-evidence links, and verification results should become support state, not prompt-local artifacts. The question is how to define algebraic and operational rules across relational, graph, textual, model-based, and human-facing operators.

\stitle{Responsibility-preserving execution.} Execution needs invariants, monitors, and repair protocols that preserve responsibility state while operators compose. It should preserve sufficient support, satisfied constraints, and actionability before an output informs action. For example, a pipeline may need to detect schema or data drift, validate or repair records, preserve conflicting evidence, or stop when constraints fail; production data/AI pipelines expose hidden dependencies and observability needs~\cite{PolyzotisSIGMOD2017ProductionML,SculleyNeurIPS2015TechnicalDebt,ShankarPVLDB2022MLObservability}, while validation and repair systems show how support can be checked and improved during execution~\cite{BreckMLSys2019DataValidation,RekatsinasPVLDB2017HoloClean,KrishnanSIGMOD2016ActiveClean}. RAIDS turns these mechanisms from controls into pipeline-level invariants: if state degrades, execution should repair, qualify, replan, or stop.

\stitle{Responsibility-aware optimization.} Optimization needs cost and admissibility models that include responsibility context. Query optimizers trade off latency, I/O, and memory; RAIDS asks when evidence coverage, provenance, privacy, fairness, compliance, carbon impact, social impact, or review burden makes a plan inadmissible or less preferred. For example, private query release may alter answers to limit disclosure~\cite{DworkTCC2006DifferentialPrivacy}, fair query processing and exposure-aware ranking may change result selection~\cite{ShetiyaICDE2022FairRangeQueries,SinghKDD2018FairExposure,ZehlikeArxiv2021FairRankingSurvey}, graph ranking and sampling may trade utility against robustness, fairness, and throughput~\cite{LiWWW2025DynamicGraphRanking,MasrourICDM2022FairGraphSampling}, and sustainable query processing may make carbon impact part of the objective~\cite{BachrasPVLDB2025SustainableDB}. Goals aligned with United Nations Sustainable Development Goals (SDGs)~\cite{UnitedNations2015SDGs} should become constraints and audit criteria in socially beneficial analytics~\cite{ding2025euleresg}. The optimization problem is admissible plan selection under responsibility context.

\stitle{Claim-level provenance and explanation.} Responsible outputs need claim/evidence graphs with decision traces~\cite{EdgeArxiv2024GraphRAG,SaadFalconNAACL2024ARES,FrielArxiv2024RAGBench}. Reports and decision outputs are collections of claims, not atomic results; each claim should carry evidence, transformations or models, policies, approvals, and unresolved uncertainty. For example, a compliance summary or scientific report should expose which records, transformations, models, and assumptions support each claim. Provenance foundations and deployment tracing show how derivations and execution traces can be queried~\cite{GreenPODS2007ProvenanceSemirings,ChapmanPVLDB2021PreprocessingProvenance,PsallidasPVLDB2023OneProvenance}; RAIDS extends this machinery from data items and jobs to claims, decisions, and action conditions. Explanation should tell the system and user what can be trusted, what remains unresolved, and what action is permitted~\cite{DoddaiahICDM2022TimeSeriesExplainability}.

\stitle{Decision and oversight protocols.} Actionability requires runtime policies for deciding what the system should do next. Human oversight must become a runtime protocol, not a passive interface after automation: support, constraint, and responsibility context should govern when to answer, qualify, ask, escalate, refuse, audit, or support contestation. For example, a public-service or health analytics system may answer when support is strong, qualify incomplete evidence, request review for high stakes, and refuse automation when safety or discrimination risks remain. Rights frameworks emphasize notice, explanation, human fallback, safety, and discrimination protection~\cite{WhiteHouse2022AIBillRights}, while risk and AI governance frameworks connect such obligations to lifecycle management~\cite{NIST2023AIRMF,EuropeanUnion2024AIAct,RajiArxiv2022OutsiderOversight}. RAIDS asks how to translate obligations into runtime modes with clear triggers, handoffs, and audits.

\stitle{Evaluation and success criteria.} Holistic evaluation shows why a single score is insufficient~\cite{LiangTMLR2023HELM,NingICDM2025TimeSeriesReasoning}; evaluation should test responsibility preservation after composition, not component quality alone. Strong components can lose support, violate constraints, or choose the wrong action mode when composed. In RAIDS, correct behavior is not always to answer: a system may be correct because it qualifies, escalates, refuses, or requests review. Success is a class of guarantees, metrics, benchmarks, and execution protocols: support, constraints, and actionability should remain meaningful after composition; policy, privacy, fairness, safety, resource, and social or environmental limits should change plans or outputs when needed; and the system should explain what it produced, why it is supported, which limits apply, and what should happen next. Progress is visible when systems preserve responsibility state across heterogeneous operators and choose the continuation mode under evidence, constraints, and oversight requirements.

\section{Conclusion}

RAIDS makes responsibility part of execution semantics: evidence, uncertainty, provenance, policy, sustainability, social impact, and oversight become state that guides computation and action. Future data systems should not only compute results; they should preserve the conditions under which results can responsibly become action.

\bibliographystyle{IEEEtran}
\bibliography{references}

\end{document}